\documentclass[12pt]{revtex4}    
\usepackage{amssymb,epsf}
\usepackage{latexsym}

\begin{document}

\title{Higher Dimensional Charged Rotating Dilaton Black Holes}
\author{A. Sheykhi$^{1,2}$\footnote{sheykhi@mail.uk.ac.ir} and  M.
Allahverdizadeh$^{1}$}
\address{$^1$Department of Physics, Shahid Bahonar University, P.O. Box 76175, Kerman, Iran\\
         $^2$Research Institute for Astronomy and Astrophysics of Maragha (RIAAM), Maragha, Iran}
\vspace*{2cm}
\begin{abstract}
\vspace*{1.5cm}\centerline{\bf Abstract} In this paper, we present
the metric for the $n$-dimensional charged slowly rotating dilaton
black hole with $N = [(n -1)/2]$ independent rotation parameters,
associated with $N$ orthogonal planes of rotation in the
background of asymptotically flat and asymptotically (anti)-de
Sitter spacetime. The mass, angular momentum and the gyromagnetic
ratio of such a black hole are determined for the arbitrary values
of the dilaton coupling constant. We find that the gyromagnetic
ratio crucially depends on the dilaton coupling constant,
$\alpha$, and decreases with increasing $\alpha$ in any dimension.

\end{abstract}
\maketitle
 \newpage

\section{Introduction}
There has been a considerable attention in general relativity in
more than four spacetime dimensions in recent years. There are
several reasons why it should be interesting to study this
extension of Einstein's theory, and in particular its black hole
solutions (for a recent review on higher dimensional black holes
see \cite{Emp1}). The first reason originates from developments in
string theory, which is believed to be the most promising approach
to quantum theory of gravity in higher dimensions. In fact, the
first successful statistical counting of black hole entropy in
string theory was performed for a five-dimensional black hole
\cite{Stro}. This example provides the best laboratory for the
microscopic string theory of black holes. Second, the AdS/CFT
correspondence relates the properties of an $n$-dimensional black
hole with those of a quantum field theory in $(n-1)$-dimensions
\cite{Aha}. According to the AdS/CFT correspondence, the rotating
black holes in AdS space are dual to certain CFTs in a rotating
space \cite{Haw}, while charged ones are dual to CFTs with
chemical potential \cite{Cham}. Furthermore, as mathematical
objects, black hole spacetimes are among the most important
Lorentzian Ricci-flat manifolds in any dimension. One striking
feature of the Einstein equations in more than four dimensions is
that many uniqueness properties holding in four dimensions are
lost. For instance, four-dimensional black holes are known to
possess a number of remarkable features, such as uniqueness,
spherical topology, dynamical stability, and the laws of black
hole mechanics. One would like to know which of these are peculiar
to four-dimensions, and which hold more generally. At least, this
study will lead to a deeper understanding of classical black holes
and of what spacetime can do at its most extreme. Besides, it was
confirmed by recent investigations that the gravity in higher
dimensions exhibits much richer dynamics than in four dimensions.
For example, the discovery of dynamical instabilities of extended
black hole horizons \cite{Gre}, and the discovery of black ring
solutions with horizons of non-spherical topology and not fully
characterized by their conserved charges \cite{Emp2}. While the
nonrotating black hole solution to the higher-dimensional
Einstein-Maxwell gravity was found several decades ago \cite{Tan},
the counterpart of the Kerr-Newman solution in higher dimensions,
that is the charged generalization of the Myers-Perry solution
\cite{Myer} in higher dimensional Einstein-Maxwell theory, still
remains to be found analytically. The most general higher
dimensional uncharged rotating black holes in anti-de Sitter space
with all rotation parameters have been recently found \cite{Gib}.
As far as we know, rotating black holes for the Maxwell field
minimally coupled to Einstein gravity in higher dimensions, do not
exist in a closed form and one has to rely on perturbative or
numerical methods to construct them in the background of
asymptotically flat \cite{kunz1,Aliev2} and AdS \cite{kunz2}
spacetime. There has also been  recent interest in constructing
the analogous charged rotating solutions in the framework of
gauged supergravity in various dimensions
\cite{Cvetic0,Cvetic1,Cvetic2}.

There has been a renewed interest in studying scalar coupled
solutions of general relativity ever since new black hole
solutions have been found in the context of string theory. The low
energy effective action of string theory contains two massless
scalars namely dilaton and axion. The dilaton field couples in a
nontrivial way to other fields such as gauge fields and results
into interesting solutions for the background spacetime
\cite{CDB1,CDB2,Hor2}. These solutions \cite{CDB1,CDB2,Hor2},
however, are all asymptotically flat. In the presence of one or
two Liouville-type dilaton potentials, black hole spacetimes which
are neither asymptotically flat nor (anti)-de Sitter [(A)dS] have
been explored by many authors (see e.g
\cite{MW,CHM,Cai,Clem,Sheykhi0,Sheykhi1}). Recently, by using the
combination of three Liouville type dilaton potentials, static
charged dilaton black hole solutions in four \cite{Gao1} and
higher dimensions \cite{Gao2} have been found in the background of
(A)dS spacetime. Such potential may arise from the
compactification of a higher dimensional supergravity model
\cite{Gid} which originates from the low energy limit of a
background string theory.

In the backdrop of the scenarios described so far it is therefore
worthwhile to study higher dimensional rotating black holes in a
spacetime with nonzero cosmological constant in the presence of
dilaton-electromagnetic coupling. For some limited values of the
dilaton coupling constant, $\alpha$, exact rotating black hole
solutions have been obtained in \cite{Fr,kun,kunz3,Bri}. For
general dilaton coupling constant, the properties of charged
rotating dilaton black holes in four
\cite{Hor1,Shi,Sheykhi2,Ghosh} and five \cite{Sheykhi3} dimensions
have been studied in the small angular momentum limit. The
generalization of these slowly rotating solutions in all higher
dimensions with a single rotation parameter have also been done
\cite{Sheykhi4,ShAll,ShAll2}. However, in more than three spatial
dimensions, black holes can rotate in different orthogonal planes,
so the general solution has several angular momentum parameters.
Indeed, an $n$-dimensional black hole can have $N = [(n -1)/2]$
independent rotation parameters, associated with $N$ orthogonal
planes of rotation where $[x]$ denotes the integer part of $x$. In
this paper we would like to extend our former works
\cite{ShAll,ShAll2} to the higher dimensional charged rotating
dilaton black holes with all rotation parameters. We then
determine the angular momentum and the gyromagnetic ratio of such
a black hole for the arbitrary values of the dilaton coupling
constant.

This paper is organized as follows. In the next section we
introduce the action and the field equations. In section
\ref{flat}, we present the metric of the higher dimensional
charged slowly rotating dilaton black hole in asymptotically flat
spacetime with all rotation parameters. We also compute the
angular momentum and the gyromagnetic ratio of the solution. In
section \ref{AdS}, we extend our solution to asymptotically (A)dS
spacetimes. The last section is devoted to summary and
conclusions.

\section{Field equations and metric}\label{flat}
Our starting point is the $n$-dimensional  theory in which gravity
is coupled to dilaton and Maxwell field with an action
\begin{eqnarray}
S &=&-\frac{1}{16\pi }\int_{\mathcal{M}} d^{n}x\sqrt{-g}\left(
R\text{ }-\frac{4}{n-2}\partial_{\mu}\Phi
\partial^{\mu}\Phi-V(\Phi
)-e^{-\frac{4\alpha \Phi}{n-2}}F_{\mu \nu }F^{\mu \nu }\right)   \nonumber \\
&&-\frac{1}{8\pi }\int_{\partial \mathcal{M}}d^{n-1}x\sqrt{-h
}\Theta (h ),  \label{Act}
\end{eqnarray}
where ${R}$ is the scalar curvature, $\Phi$ is the dilaton field,
$F_{\mu \nu }=\partial _{\mu }A_{\nu }-\partial _{\nu }A_{\mu }$
is the electromagnetic field tensor, and $A_{\mu }$ is the
electromagnetic potential. $\alpha $ is an arbitrary constant
governing the strength of the coupling between the dilaton and the
Maxwell field and $V(\Phi )$ is the potential for $\Phi$. The last
term in Eq. (\ref{Act}) is the Gibbons-Hawking surface term. It is
required for the variational principle to be well-defined. The
factor $\Theta$ represents the trace of the extrinsic curvature
for the boundary ${\partial \mathcal{M}}$ and $h$ is the induced
metric on the boundary. While $\alpha=0$ corresponds to the usual
Einstein-Maxwell-scalar theory, $\alpha=1$ indicates the
dilaton-electromagnetic coupling that appears in the low energy
string action in Einstein's frame.

Varying the action (\ref{Act}) with respect to the gravitational
field $g_{\mu \nu }$, the dilaton field $\Phi $ and the gauge
field $A_{\mu }$, we obtain the field equations
\begin{equation}
R_{\mu \nu }=\frac{4}{n-2} \left(\partial _{\mu }\Phi
\partial _{\nu }\Phi+\frac{1}{4}g_{\mu \nu }V(\Phi )\right)+2e^{\frac{-4\alpha \Phi}{n-2}}\left( F_{\mu \eta }F_{\nu }^{\text{
}\eta }-\frac{1}{2(n-2)}g_{\mu \nu }F_{\lambda \eta }F^{\lambda
\eta }\right) ,  \label{FE1}
\end{equation}
\begin{equation}
\nabla ^{2}\Phi =\frac{n-2}{8}\frac{\partial V}{\partial \Phi
}-\frac{\alpha }{2}e^{\frac{-4\alpha \Phi}{n-2}}F_{\lambda \eta
}F^{\lambda \eta },  \label{FE2}
\end{equation}
\begin{equation}
\partial_{\mu}{\left(\sqrt{-g} e^{\frac{-4\alpha \Phi}{n-2}}F^{\mu \nu }\right)}=0. \label{FE3}
\end{equation}
We would like to find the rotating solutions of the above field
equations with all rotation parameters in all higher dimensions in
the limit of slow rotation. The rotation group in $n$-dimensions
is $SO(n-1)$ and therefore the number of independent rotation
parameters for a localized object is equal to the number of
Casimir operators, which is $N = [(n -1)/2]$, where $[x]$ is the
integer part of $x$. For small rotation parameters, we can solve
Eqs. (\ref{FE1})-(\ref{FE3}) to first order in the angular
momentum parameters $a_{{i}}$. Inspection of the Myers-Perry
solution \cite{Myer} shows that the terms in the metric that
change to the first order of the rotational parameters $a_{i}$'s
are $g_{t\phi_{i}}$'s $(i=1...N)$. Similarly, the dilaton field
does not change to $O(a_{i})$ and $A_{\phi_{i}}$'s are the
components of the vector potential that change. Therefore,
inspired by the Myers-Perry metric \cite{Myer}, for infinitesimal
angular momentum, we assume the metric being of the following form
\begin{eqnarray}\label{metric}
ds^2 &=&-U(r)dt^2+{dr^2\over W(r)}- 2  f(r)d{t}\sum
_{i=1}^{N}a_{{i}}\mu_{{i}}^{2}
d{\phi_{{i}}}\nonumber \\
 &&+ r^2 R^2(r)\sum _{i=1}^{N}\left(d{\mu_{{i}}}^2 +\mu_{{i}}^{2}
d{\phi_{{i}}}^{2}\right)+ \varepsilon  r^2 R^2(r) d{\nu}^2,
\end{eqnarray}
Where for even $n$ we have $N=\frac{n-2}{2}$ and $\varepsilon=1$,
whereas for odd n, $N=\frac{n-1}{2}$ and $\varepsilon=0$. The
$\mu_{{i}}$ (and $\nu$ , for even dimensional cases) coordinates
are not independent but have to obey the constraint
\begin{equation}
\sum _{i=1}^{N}\mu_{{i}}^{2}+\varepsilon \nu^2=1.\label{sum}
\end{equation}
The functions $U(r)$, $W(r)$, $R(r)$ and $f(r)$ should be
determined. For small $a_{i}$ $(i=1...N)$, we can expect to have
solutions with $U(r)$ and $W(r)$ still  functions of $r$ alone.
The $t$ component of the Maxwell equations can be integrated
immediately to give
\begin{equation}\label{Ftr}
F_{tr}=\sqrt{\frac{U(r)}{W(r)}}\frac{Q e^{\frac{4\alpha
\Phi}{n-2}}}{\left( rR\right) ^{n-2}} ,
\end{equation}
where $Q$, an integration constant, is the electric charge of the
black hole. In general, in the presence of rotation, we have
$A_{\phi_{i}}\neq0$. In addition we can write
\begin{equation}\label{Aphi}
 A_{\phi_{i}}=- Q C(r)a_{{i}}\mu_{{i}}^{2}   \hspace{0.5cm}{\text{(no sum on $i$)}}.
\end{equation}

\section{Rotating Dilaton Black Holes in flat spacetime}\label{flat}
\begin{figure}[tbp]
\epsfxsize=7cm \centerline{\epsffile{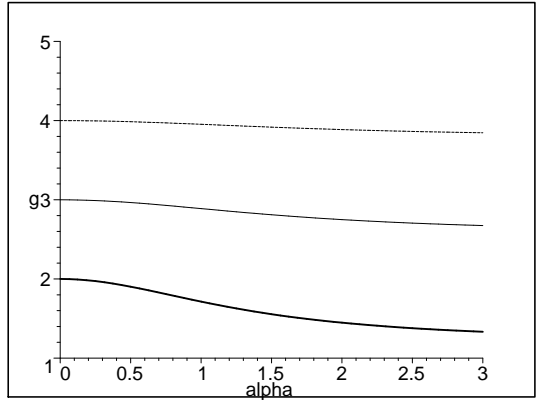}} \caption{The
behaviour of the gyromagnetic ratio $g$ versus $\protect\alpha$ in
various dimensions for $r_{-}=1$, $r_{+}=2$. $n=4$ (bold line),
$n=5$ (continuous line), and $n=6$ (dashed line).} \label{figure1}
\end{figure}

We begin by looking for the asymptotically flat solutions,
therefore we set $V(\Phi)=0$ in the field equations. In a recent
work \cite{ShAll}, we found a class of asymptotically flat slowly
rotating charged dilaton black hole solution in higher dimensions
with single rotation parameter. Here we are looking for the
asymptotically flat slowly rotating dilaton black hole solutions
with all rotation parameters in all higher dimensions. Inserting
the metric (\ref{metric}), the Maxwell fields (\ref{Ftr}) and
(\ref{Aphi}) into the field equations (\ref{FE1})-(\ref{FE3}), one
can show that the static part of the metric leads to the following
solutions \cite{Hor2}
\begin{eqnarray}\label{U}
U(r)&=&\left[1-\left(\frac{r_{+}}{r}\right)^{n-3}\right]\left[1-\left(\frac{r_{-}}{r}\right)^{n-3}\right]^{1-\gamma\left(n-3\right)}
,
\end{eqnarray}
\begin{eqnarray}\label{W}
W(r)&=&\left[1-\left(\frac{r_{+}}{r}\right)^{n-3}\right]\left[1-\left(\frac{r_{-}}{r}\right)^{n-3}\right]^{1-\gamma},
\end{eqnarray}
\begin{eqnarray}\label{R}
R(r)&=&\left[1-\left(\frac{r_{-}}{r}\right)^{n-3}\right]^{\gamma/2},
\end{eqnarray}
\begin{eqnarray}\label{Phi}
\Phi(r)=\frac{n-2}{4}\sqrt{\gamma(2+3\gamma-n\gamma)}\ln
\left[1-\left(\frac{r_{-}}{r}\right)^{n-3}\right],
\end{eqnarray}
while the rotating part of the metric admits a solution
\begin{eqnarray}\label{f0}
f(r)&=&\left(n-3\right)\left(\frac{r_{+}}{r}\right)^{n-3}\left[1-\left(\frac{r_{-}}{r}\right)^{n-3}\right]^{\frac{n-3-\alpha^{2}}
{n-3+\alpha^{2}}}\nonumber\\
&&+\frac{(\alpha^{2}-n+1)(n-3)^{2}}{\alpha^{2}+n-3}r_{-}^{n-3}r^{2}\left[1-\left(\frac{r_{-}}{r}\right)^{n-3}\right]
^{\gamma}\nonumber\\
&& \times \int
\left[1-\left(\frac{r_{-}}{r}\right)^{n-3}\right]^{\gamma(2-n)}
\frac{dr}{r^{n}},
\end{eqnarray}
\begin{eqnarray}\label{C}
 C(r)= \frac{1}{r^{n-3}}.
\end{eqnarray}
We can also perform the integration and express the solution in
terms of hypergeometric function
\begin{eqnarray}\label{f}
f(r)&=&\left(n-3\right)\left(\frac{r_{+}}{r}\right)^{n-3}\left[1-\left(\frac{r_{-}}{r}\right)^{n-3}\right]^{\frac{n-3-\alpha^{2}}
{n-3+\alpha^{2}}}\nonumber\\
&&+\frac{(\alpha^{2}-n+1)(n-3)^{2}}{(1-n
)(\alpha^{2}+n-3)}(\frac{r_{-}}{r})^{n-3}\left[1-\left(\frac{r_{-}}{r}\right)^{n-3}\right]
^{\gamma}\nonumber\\
&& \times  _{2}F_{1} \left(\left[(n-2)\gamma,\frac
{n-1}{n-3}\right],\left[\frac {2n-4}{n-3}\right],\left({\frac
{r_{-}}{r}}\right) ^{n-3} \right).
\end{eqnarray}
Here $r_+$ and $r_{-}$ are the event horizon and Cauchy horizon of
the black hole, respectively. The constant $\gamma$ is
\begin{equation}\label{gamma}
\gamma=\frac{2\alpha^{2}}{(n-3)(n-3+\alpha^{2})}.
\end{equation}
The charge $Q$ is related to $r_+$ and $r_{-}$ by
\begin{equation}\label{Q}
Q^{2}=\frac{(n-2)(n-3)^{2}}{2(n-3+\alpha^{2})}r_{+}^{n-3}r_{-}^{n-3},
\end{equation}
and the physical mass of the black hole is obtained as follows
\cite{Fang}
\begin{equation}\label{mass}
{M}=\frac{\Omega
_{n-2}}{16\pi}\left[(n-2)r^{n-3}_{+}+\frac{n-2-p(n-4)}{p+1}r^{n-3}_{-}\right],
\end{equation}
where $\Omega _{n-2}$ denotes the area of the unit $(n-2)$-sphere
and the constant $p$ is
\begin{equation}\label{pp}
{p}=\frac{(2-n)\gamma}{(n-2)\gamma-2}.
\end{equation}
The metric corresponding to (\ref{U})-(\ref{f}) is asymptotically
flat. In the special case $n=4$, the static part of our solution
reduces to
\begin{equation}\label{U4}
U(r)= W(r)= \left( 1-{\frac {r_{+}}{r}} \right) \left( 1-{\frac
{r_{-}}{r}} \right) ^{\,{\frac {{1-\alpha}^{2}}{1+{\alpha}^{2}}}},
\end{equation}
\begin{equation}\label{R4}
R \left( r \right) = \left( 1-{\frac {r_{-}}{r}} \right) ^{{\frac
{{\alpha }^{2}}{1+{\alpha}^{2}}}},
\end{equation}
\begin{equation}\label{Phi4}
\Phi \left( r \right) =\frac{\alpha}{\left( 1+{\alpha}^{2}
\right)}\ln  \left(1- {\frac {r_{-}}{r}} \right),
\end{equation}
while the rotating part reduces to
\begin{equation}
\label{fhor} f(r)=\frac{r^{2}(1+\alpha^{2})^{2}
(1-\frac{r_{-}}{r})^{\frac{2\alpha^{2}}{1+\alpha^{2}}}}{(1-\alpha^{2})
(1-3\alpha^{2})r^{2}_{-}}-\left(1-\frac{r_{-}}{r}\right)^{\frac{1-\alpha^{2}}
{1+\alpha^{2}}}\left(1+\frac{(1+\alpha^{2})^{2}r^{2}}{(1-\alpha^{2})(1-3\alpha^{2})
r^{2}_{-}}+\frac{(1+\alpha^{2})r}{(1-\alpha^{2})r_{-}}-\frac{r_{+}}{r}\right),
\end{equation}
which is the four-dimensional charged slowly rotating dilaton
black hole solution of Horne and Horowitz \cite{Hor1}. One may
also note that in the absence of a non-trivial dilaton
($\alpha=0=\gamma$), our solutions reduce to
\begin{equation}\label{U0}
U \left( r \right) = W(r)= \left[ 1- \left( {\frac {r_{+}}{r}}
\right) ^{n-3}
 \right]  \left[ 1- \left( {\frac {r_{-}}{r}} \right) ^{n-3}
 \right],
\end{equation}
\begin{equation}\label{f0}
f \left( r \right)
=(n-3)\left[\frac{r^{n-3}_{-}+r^{n-3}_{+}}{r^{n-3}}-\left(\frac{r_{+}r_{-}}{r^{2}}\right)^{n-3}\right],
\end{equation}
which describe $n$-dimensional Kerr-Newman black hole in the limit
of slow rotation \cite{Aliev2}.
Next, we calculate the angular momentum and the gyromagnetic ratio
of these rotating dilaton black holes which appear in the limit of
slow rotation parameters. The angular momentum of the dilaton black
hole can be calculated through the use of the quasi-local
formalism of the Brown and York \cite{BY}. According to the
quasilocal formalism, the quantities can be constructed from the
information that exists on the boundary of a gravitating system
alone. Such quasilocal quantities will represent information about
the spacetime contained within the system boundary, just like the
Gauss's law. In our case the finite stress-energy tensor can be
written as
\begin{equation}
T^{ab}=\frac{1}{8\pi }\left(\Theta^{ab}-\Theta h ^{ab}\right) ,
\label{Stres}
\end{equation}
which is obtained by variation of the action (\ref{Act}) with
respect to the boundary metric $h _{ab}$. To compute the
angular momentum of the spacetime, one should choose a spacelike surface $%
\mathcal{B}$ in $\partial \mathcal{M}$ with metric $\sigma _{ij}$,
and write the boundary metric in ADM form
\[
\gamma _{ab}dx^{a}dx^{a}=-N^{2}dt^{2}+\sigma _{ij}\left( d\varphi
^{i}+V^{i}dt\right) \left( d\varphi ^{j}+V^{j}dt\right) ,
\]
where the coordinates $\varphi ^{i}$ are the angular variables
parameterizing the hypersurface of constant $r$ around the origin,
and $N$ and $V^{i}$ are the lapse and shift functions
respectively. When there is a Killing vector field $\mathcal{\xi
}$ on the boundary, then the quasilocal conserved quantities
associated with the stress tensors of Eq. (\ref{Stres}) can be
written as
\begin{equation}
Q(\mathcal{\xi )}=\int_{\mathcal{B}}d^{n-2}\varphi \sqrt{\sigma }T_{ab}n^{a}%
\mathcal{\xi }^{b},  \label{charge}
\end{equation}
where $\sigma $ is the determinant of the metric $\sigma _{ij}$, $\mathcal{%
\xi }$ and $n^{a}$ are, respectively, the Killing vector field and
the unit normal vector on the boundary $\mathcal{B}$. For
boundaries with rotational ($\varsigma =\partial /\partial \varphi
$) Killing vector field, we can write the corresponding quasilocal
angular momentum as follows
\begin{eqnarray}
J &=&\int_{\mathcal{B}}d^{n-2}\varphi \sqrt{\sigma
}T_{ab}n^{a}\varsigma ^{b},  \label{Angtot}
\end{eqnarray}
provided the surface $\mathcal{B}$ contains the orbits of
$\varsigma $. Finally, the angular momentum of the black holes can
be calculated by using Eq. (\ref{Angtot}). We find
\begin{equation}
{{J}_{{i}}}=\frac{a_{{i}}\Omega
_{n-2}}{8\pi}\left(r^{n-3}_{+}+\frac{(n-3)(n-1-\alpha^{2})r^{n-3}_{-}}{(n-3+\alpha^{2})(n-1)}\right).
\label{JJ}
\end{equation}
For $a_{i}=0$, the angular momentum vanishes, and therefore
$a_{i}$'s are the rotational parameters of the dilaton black hole.
For $n=4$, we have only one rotation parameter, $a$. In this case,
the corresponding angular momentum reduces to
\begin{equation}
{{J}}=\frac{a}{2}\left(r_{+}+\frac{3-\alpha^2}{3(1+\alpha^2)}r_{-}\right),
\label{J4}
\end{equation}
which restores the angular momentum of the four-dimensional Horne
and Horowitz solution \cite{Hor1}, while in the absence of dilaton
field $(\alpha=0)$, the angular momentum reduces to
\begin{equation}
{{J}_{{i}}}=\frac{a_{i}\Omega
_{n-2}}{8\pi}\left(r^{n-3}_{+}+r^{n-3}\right), \label{Jn}
\end{equation}
which is the angular momentum  of the $n$-dimensional Kerr-Newman
black hole. Next, we are going to calculate the gyromagnetic ratio
of this rotating dilaton black holes. As we know, the gyromagnetic
ratio is an important characteristic of the Kerr-Newman-AdS black
hole. Indeed, one of the remarkable facts about a Kerr-Newman
black hole in asymptotically flat spacetime is that it can be
assigned a gyromagnetic ratio $g = 2$, just as an electron in the
Dirac theory. It should be noted that, unlike four dimensions, the
value of the gyromagnetic ratio is not universal in higher
dimensions \cite{Aliev3}. Besides, scalar fields such as the
dilaton,  modify the value of the gyromagnetic ratio of the black
hole and consequently it does not possess the gyromagnetic ratio
$g = 2$ of the Kerr-Newman black hole \cite{Hor1}.The magnetic
dipole moments for this asymptotically flat slowly rotating
dilaton black hole can be defined as
\begin{equation}\label{mu}
{\mu_{{i}}}=Qa_{{i}}.
\end{equation}
The gyromagnetic ratio is defined as a constant of proportionality
in the equation for the magnetic dipole moments
\begin{equation}\label{mu}
{\mu_{{i}}}=g\frac{QJ_{{i}}}{2M}.
\end{equation}
 Substituting $M$ and $J_{{i}}$ from Eqs. (\ref{mass}) and  (\ref{JJ}),
 the gyromagnetic ratio $g$ can be obtained as
\begin{equation}\label{g}
g=\frac{(n-1)(n-2)[(n-3+\alpha^2)r^{n-3}_{+}+(n-3-\alpha^2)r^{n-3}_{-}]}{
(n-1)(n-3+\alpha^2)r^{n-3}_{+}+(n-3)(n-1-\alpha^2)r^{n-3}_{-}}.
\end{equation}
The above expression shows that the value of the gyromagnetic
ratio $g$ is the same as in the case of single rotation parameter
\cite{ShAll}. We can also see that the dilaton field modifies the
value of the gyromagnetic ratio through the dilaton coupling
constant $\alpha$ which measures the strength of the
dilaton-electromagnetic coupling. This is in agreement with the
arguments in \cite{Hor1,ShAll}. We have shown the behaviour of the
gyromagnetic ratio $g$ of the dilatonic black hole versus
$\protect\alpha$ in Fig. \ref{figure1}. From this figure we find
out that the gyromagnetic ratio decreases with increasing $\alpha$
in any dimension. In the absence of a non-trivial dilaton
$(\alpha=0)$, the gyromagnetic ratio reduces to
\begin{equation}\label{gkerr-newman}
{g}=n-2,
\end{equation}
which is the gyromagnetic ratio of the $n$-dimensional Kerr-Newman
black hole in the slow rotation limit \cite{Aliev2}. When $n=4$,
Eq. (\ref{g}) reduces to
\begin{equation}\label{gHor}
{g}=2-\frac{4\alpha^{2}r_{-}}{(3-\alpha^{2})r_{-}+3(1+\alpha^{2})r_{+}},
\end{equation}
which is the gyromagnetic ratio of the four-dimensional Horne and
Horowitz dilaton black hole \cite{Hor1}.

\section{Rotating Dilaton Black Holes in AdS spacetime} \label{AdS}
Now we consider the solutions of Eqs. (\ref{FE1})-(\ref{FE3}) with
Liouville-type dilaton potential. For arbitrary value of $\alpha $
in (A)dS  space the form of the dilaton potential in arbitrary
dimensions is chosen as \cite{Gao2}
\begin{eqnarray}\label{V1}
V(\Phi)&=&\frac{\Lambda}{3(n-3+\alpha^2)^{2}}\left[-\alpha^2(n-2)\left(n^{2}-n\alpha^{2}-6n+\alpha^{2}+9\right)
e^{\frac{-4(n-3)\Phi}{(n-2)\alpha}}\right. \nonumber
\\
&& \left.+(n-2)(n-3)^{2}(n-1-\alpha^{2})
e^{\frac{4\alpha\Phi}{n-2}}+4\alpha^{2}(n-3)(n-2)^{2}
e^{\frac{-2\Phi(n-3-\alpha^{2})}{(n-2)\alpha}}\right].
\end{eqnarray}
Here $\Lambda $ is the cosmological constant. It is clear the
cosmological constant is coupled to the dilaton in a very
nontrivial way. This type of the dilaton potential was introduced
for the first time by Gao and Zhang \cite{Gao1,Gao2}. They
derived, by applying a coordinates transformation which recast the
solution in the Schwarzchild coordinates system, the static
dilaton black hole solutions in the background of (A)dS universe.
For this purpose, they required the existence of the (A)dS dilaton
black hole solutions and extracted successfully the form of the
dilaton potential leading to (A)dS-like solutions. They also
argued that this type of derived potential can be obtained when a
higher dimensional theory is compactified to four dimensions,
including various supergravity models \cite{Gid}. In the absence
of the dilaton field the action (\ref{Act}) reduces to the action
of Einstein-Maxwell gravity with cosmological constant. Inserting
the metric (\ref{metric}), the Maxwell fields (\ref{Ftr}) and
(\ref{Aphi}), and the dilaton potential (\ref{V1}) into the field
equations (\ref{FE1})-(\ref{FE3}), one can show that the static
part of the metric leads to the following solutions \cite{Gao2}
\begin{eqnarray}\label{U1}
U(r)&=&\left[1-\left(\frac{r_{+}}{r}\right)^{n-3}\right]\left[1-\left(\frac{r_{-}}{r}\right)^{n-3}\right]^{1-\gamma\left(n-3\right)}-\frac{1}{3}\Lambda
r^2\left[1-\left(\frac{r_{-}}{r}\right)^{n-3}\right]^{\gamma} ,
\end{eqnarray}
\begin{eqnarray}\label{W1}
W(r)&=&\Bigg{\{}\left[1-\left(\frac{r_{+}}{r}\right)^{n-3}\right]
\left[1-\left(\frac{r_{-}}{r}\right)^{n-3}\right]^{1-\gamma\left(n-3\right)}-\frac{1}{3}\Lambda
r^2\left[1-\left(\frac{r_{-}}{r}\right)^{n-3}\right]^{\gamma}\Bigg
{\}}\nonumber
\\&& \times
\left[1-\left(\frac{r_{-}}{r}\right)^{n-3}\right]^{\gamma(n-4)},
\end{eqnarray}
\begin{eqnarray}\label{Phi1}
\Phi(r)&=&\frac{n-2}{4}\sqrt{\gamma(2+3\gamma-n\gamma)}\ln\left[1-\left(\frac{r_{-}}{r}\right)^{n-3}\right],
\end{eqnarray}
\begin{eqnarray}\label{R1}
R(r)&=&\left[1-\left(\frac{r_{-}}{r}\right)^{n-3}\right]^{\gamma/2},
\end{eqnarray}
while we obtain the following solution for the rotating part of
the metric
\begin{eqnarray}\label{f1}
f(r)&=&\frac{2\Lambda r^2}{(n-1)(n-2)}\left[1- \left({\frac
{r_{-}}{r}} \right) ^{n-3} \right]^{
\gamma}+\left(n-3\right)\left(\frac{r_{+}}{r}\right)^{n-3}\left[1-\left(\frac{r_{-}}{r}\right)^{n-3}\right]^{\frac{n-3-\alpha^{2}}
{n-3+\alpha^{2}}}\nonumber\\
&&+\frac{(\alpha^{2}-n+1)(n-3)^{2}}{\alpha^{2}+n-3}r_{-}^{n-3}r^{2}\left[1-\left(\frac{r_{-}}{r}\right)^{n-3}\right]
^{\gamma} \times \int
\left[1-\left(\frac{r_{-}}{r}\right)^{n-3}\right]^{\gamma(2-n)}
\frac{dr}{r^{n}},
\end{eqnarray}
\begin{eqnarray}\label{C1}
 C(r)= \frac{1}{r^{n-3}}.
\end{eqnarray}
We can also perform the integration and express the solution in
terms of hypergeometric function
\begin{eqnarray}\label{f2}
f(r)&=&\frac{2\Lambda r^2}{(n-1)(n-2)}\left[1- \left({\frac
{r_{-}}{r}} \right) ^{n-3} \right]^{
\gamma}+\left(n-3\right)\left(\frac{r_{+}}{r}\right)^{n-3}\left[1-\left(\frac{r_{-}}{r}\right)^{n-3}\right]^{\frac{n-3-\alpha^{2}}
{n-3+\alpha^{2}}}\nonumber\\
&&+\frac{(\alpha^{2}-n+1)(n-3)^{2}}{(1-n
)(\alpha^{2}+n-3)}(\frac{r_{-}}{r})^{n-3}\left[1-\left(\frac{r_{-}}{r}\right)^{n-3}\right]
^{\gamma}\nonumber\\
&& \times  _{2}F_{1} \left(\left[(n-2)\gamma,\frac
{n-1}{n-3}\right],\left[\frac {2n-4}{n-3}\right],\left({\frac
{r_{-}}{r}}\right) ^{n-3} \right).
\end{eqnarray}
It is apparent that the metric corresponding to
(\ref{U1})-(\ref{f2}) is asymptotically (A)dS. For $\Lambda=0$,
the above solutions recover our previous results for
asymptotically flat rotating dilaton black holes. It is worth
noting that in the absence of a non-trivial dilaton
($\alpha=0=\gamma $), our solutions reduce to
\begin{equation}\label{U00}
U \left( r \right) = W(r)= \left[ 1- \left( {\frac {r_{+}}{r}}
\right) ^{n-3}
 \right]  \left[ 1- \left( {\frac {r_{-}}{r}} \right) ^{n-3}
 \right]-\frac{1}{3}\Lambda r^2
\end{equation}
\begin{equation}\label{f00}
f \left( r \right)
=(n-3)\left[\frac{r^{n-3}_{-}+r^{n-3}_{+}}{r^{n-3}}-\left(\frac{r_{+}r_{-}}{r^{2}}\right)^{n-3}\right]+{\frac
{2\Lambda\,{r}^{2}}{ \left( n-1 \right)  \left( n-2 \right) }} ,
\end{equation}
which describe $n$-dimensional charged Kerr-(A)dS black hole in
the limit of slow rotation. We can also calculate the angular
momentum and the gyromagnetic ratio of these asymptotically (A)dS
rotating dilaton black holes. We find
\begin{equation}
{{J}_{{i}}}=\frac{a_{{i}}\Omega
_{n-2}}{8\pi}\left(r^{n-3}_{+}+\frac{(n-3)(n-1-\alpha^{2})r^{n-3}_{-}}{(n-3+\alpha^{2})(n-1)}\right).
\label{J1}
\end{equation}
 \begin{equation}\label{g1}
g=\frac{(n-1)(n-2)[(n-3+\alpha^2)r^{n-3}_{+}+(n-3-\alpha^2)r^{n-3}_{-}]}{
(n-1)(n-3+\alpha^2)r^{n-3}_{+}+(n-3)(n-1-\alpha^2)r^{n-3}_{-}}.
\end{equation}
One can see that in the linear approximation in the rotational
parameters $a_{i}$'s, the above expressions for ${J}_{{i}}$ and
$g$ turns out to be same as those obtained in Eqs. (\ref{Jn}) and
(\ref{g}) for asymptotically flat slowly rotating dilaton black
hole. This means that the dilaton potential (cosmological constant
term in dilaton gravity) does not change the angular momentum and
the gyromagnetic ratio of the rotating (A)dS dilaton black holes,
as pointed out in \cite{Ghosh}.

\section{Summary and Conclusion}\label{sum}
To summarize, we have presented the metric for the $n$-dimensional
charged slowly rotating dilaton black hole with $N = [(n -1)/2]$
independent rotation parameters, associated with $N$ orthogonal
planes of rotation. These solutions are asymptotically flat and
(anti)-de Sitter. We started from the nonrotating charged dilaton
black hole solutions in $n$-dimensions. Then, we considered the
effect of adding $N$ small rotational parameters $a_{{i}}$'s to
the solution and then successfully obtained the solution for such
a black hole in all higher dimensions. We discarded any terms
involving $a_{{i}}^2$ or higher powers in $a_{{i}}$. For small
rotational parameters, the terms in the metric which change are
$g_{t\phi_{i}}$'s. The vector potential is chosen to have a
nonradial component $A_{\phi_{i}} = - Q C(r)a_{{i}}\mu_{{i}}^{2}$.
to represent the magnetic field due to the rotation of the black
hole. We have obtained the angular momentum and the gyromagnetic
ratio which appear up to the linear order of the  rotational
parameters $a_{{i}}$'s. We have shown that the gyromagnetic ratio
crucially depends on the dilaton coupling constant, $\alpha$, and
decreases with increasing $\alpha$ in any dimension.

\acknowledgments{This work has been supported financially by
Research Institute for Astronomy and Astrophysics of Maragha,
Iran.}


\end{document}